\documentclass[pra,aps,twocolumn,showpacs]{revtex4}
\usepackage{amssymb}
\usepackage{amsmath}
\usepackage[dvips]{graphicx}
\usepackage[english]{babel}

\begin{document}

\title{Effect of thermal fluctuations on spin degrees of freedom in spinor
Bose-Einstein condensates}
\author{W. V. Pogosov and K. Machida}
\affiliation{Department of Physics, Okayama University, Okayama 700-8530, Japan}
\date{\today }

\begin{abstract}
We consider the effect of thermal fluctuations on rotating spinor $F=1$
condensates in axially-symmetric vortex phases, when all the three hyperfine
states are populated. We show that the relative phase among different
components of the order parameter can fluctuate strongly due to the weakness
of the interaction in the spin channel. These fluctuations can be
significant even at low temperatures. Fluctuations of relative phase lead to
significant fluctuations of the local transverse magnetization of the
condensate. We demonstrate that these fluctuations are much more pronounced
for the antiferromagnetic state than for the ferromagnetic one.
\end{abstract}

\maketitle

\section{Introduction}

Properties of rotating spinor Bose-Einstein condensates attract a lot of
attention now. First examples of these systems with hyperfine spin $F=1$
were found in optically trapped $^{23}$Na \cite{Stenger}. Vortex phase
diagram of spinor condensates is very rich, since the order parameter has
three components in $F=1$ case and five components in $F=2$ case.
Topological excitations in spinor condensates were studied theoretically in
a large number of articles see, e.g., Refs \cite%
{Mizushima1,Isoshima1,Reijnders,Ueda1,Pogosov}.

At the same time, an interest is now growing to temperature effects in
atomic condensates. Refs. \cite{Trombettoni,Simula,Baym,Simula2,Hutchinson}
study theoretically the Berezinskii-Kosterlitz-Thouless (BKT) transition
associated with the proliferation of thermally-excited vortex-antivortex
pairs. For instance, in Ref. \cite{Simula} it was shown that in quasi
two-dimensional condensates BKT transition can occur at rather low
temperatures, $T\sim 0.5T_{c}$, at number of particles in the system $N\sim
10^{4}$. Recently, some signatures of possible BKT phase were also found
close to the critical temperature $T_{c}$ in experimental work \cite%
{BKTexper}, where condensates in optical lattice have been studied. 
Finally, experimental evidence for the BKT transition in trapped condensates
was reported in Ref. \cite{Dalibard}. 
Refs. 
\cite{Castin,Pogos} deal with the thermal fluctuations of positions of
vortices in rotated scalar condensates. Note that, according to the
Mermin-Wagner-Hohenberg theorem, Bose-Einstein condensation is not possible
in 2D homogeneous systems. However, application of the trapping potential
leads to the macroscopic occupation of the ground state of Bose gas.

The aim of the present paper is to study the effect of thermal fluctuations
in rotated quasi two-dimensional \textit{spinor} condensates. These systems
have a specific degree of freedom, associated with the relative angle among
different components of the order parameter corresponding to different
hyperfine state. In other words, this angle determines coherence among
components of the order parameter. Also it influences a transverse
magnetization of the condensate. In this paper, we focus on thermal
fluctuations of this angle. Note that experimentally, at present time, it is
possible to study the condensate phase \cite{Wang,Wheeler,Saba}, see also
Ref \cite{Chapman}. In addition, recently, a new and nondestructive method
for measuring the local magnetization of the condensate was proposed and
successfully applied in Ref. \cite{Higbie}.

We show that the relative angle among hyperfine components of the order
parameter in 2D case can experience strong thermal fluctuation even at low
temperatures. The reason is the weakness of the spin energy of the system as
compared to interactions in density channel. Also fluctuations of this angle
lead to significant relative fluctuations of the local transverse
magnetization of the condensate, which are much larger in the
antiferromagnetic case than in the ferromagnetic one.

This paper is organized as follows. In Section II, we give a basic
formulation of the problem. In Section II, we discuss our main results for
the fluctuations of angle and spin textures. We conclude in Section III.

\section{Basic formulation}

We consider harmonically-trapped quasi 2D Bose-Einstein condensate with spin 
$F=1$. The trapping potential is given by

\begin{equation}
U(r)=\frac{m\omega _{\perp }^{2}r^{2}}{2},  \label{1}
\end{equation}%
where $\omega _{\perp }$ is a trapping frequency, $m$ is the mass of the
atom, and $r$ is the radial coordinate. The system is rotated with the
angular velocity $\mathbf{\Omega }$, well below the critical rotation speed $%
\omega _{\perp }$, and the number of atoms in the cloud is $N$. In this
paper, we restrict ourselves on the range of temperatures much smaller than $%
T_{c}$. Therefore, we can neglect a noncondensate contribution to the free
energy of the cloud. The total energy of the system in this approximation
coincides with the energy of the condensate. For the number of condensed
particles, we use the ideal gas result:%
\begin{equation}
N(T)=N\left[ 1-\left( \frac{T}{T_{c}}\right) ^{2}\right] .  \label{2}
\end{equation}%
At the same time,

\begin{equation}
\frac{\hbar \omega _{\perp }}{k_{B}T_{c}}=\sqrt{\frac{\zeta (2)}{N}},
\label{3}
\end{equation}%
where $\zeta (2)$\ is a Riemann zeta function, $\sqrt{\zeta (2)}\approx 1.28$%
. Eqs. (2) and (3) remain accurate even for the case of interacting
particles \cite{Gies}. We also introduce a dimensionless temperature $%
t=T/T_{c}$. Since we are considering low temperatures, $T\sim 0.1T_{c}$,
temperature dependence of condensed particles number can be neglected, $%
N(T)\simeq N$.

The order parameter in the $F=1$ condensate has three components $\Psi _{j}$ 
$(j=-1,0,1)$. The free energy of the system can be written as \cite%
{Machida,Ho}%
\begin{eqnarray}
F &=&\hbar \omega _{\perp }N\int dS\left[ \Psi _{j}^{\ast }\widehat{h}\Psi
_{j}+2\pi g_{n}\Psi _{j}^{\ast }\Psi _{k}^{\ast }\Psi _{j}\Psi _{k}\right. 
\notag \\
&&+2\pi g_{s}\Psi _{j}^{\ast }\Psi _{l}^{\ast }(F_{a})_{jk}(F_{a})_{lm}\Psi
_{k}\Psi _{m}  \notag \\
&&\left. -i\mathbf{\Omega \cdot }\Psi _{j}^{\ast }(\mathbf{\nabla \times r}%
)\Psi _{j}\right] ,  \label{4}
\end{eqnarray}%
where the integration is performed over the system area, repeated indices
are summed, $F_{a}$ $(a=x,y,z)$ is the angular momentum operator, which can
be expressed in a matrix form through the usual Pauli matrices, $\widehat{h}$
is the one-body Hamiltonian, given by%
\begin{equation}
\widehat{h}=-\frac{\nabla ^{2}}{2}+\frac{r^{2}}{2}.  \label{5}
\end{equation}

Constants $g_{n}$ and $g_{s\bigskip }$ characterize interactions in density
and spin channels and are given by

\begin{equation}
g_{n}=\frac{(a_{0}+2a_{2})n_{z}}{3},  \label{6}
\end{equation}

\begin{equation}
g_{s}=\frac{(a_{2}-a_{0})n_{z}}{3},  \label{7}
\end{equation}%
where $a_{0}$ and $a_{2}$ are scattering lengths for atoms with total spin 0
and 2, and $n_{z}$ is the concentration of atoms in longitudinal direction.
In real spinor condensates, $\left\vert g_{s}\right\vert \ll \left\vert
g_{n}\right\vert $, since $a_{0}\approx a_{2}$. Typically, $\left\vert
g_{s}/g_{n}\right\vert \sim 0.001-0.01$, and this ratio can be tuned. In
this paper, we study the case of relatively dilute condensate and take $%
g_{n}=10$. We will consider different values of $N$ but at fixed value of
interaction parameter $g_{n}$. This is possible, since, in the case of a
single layer cloud, we can always tune the trapping frequency in the
longitudinal direction keeping $g_{n}$\ constant. To ensure the regime of
quasi-two-dimensionality, we can also tune $\omega _{\perp }$. In this case,
we have to change the rotation speed to keep the dimensionless rotation
speed the same, and the themperature to fix dimensionless $t$. In real
atomic condensates, $a$ is approximately several nanometers. The most
realistic value of $N$ for this $g_{n}$ is close to $10^{3}$, and to
illustrate the effect of $N$ we will consider the following range: $%
10^{2}\lesssim N\lesssim 10^{4}$.

The total magnetization of the condensate is fixed:

\begin{equation}
M=\int dS\left\vert \Psi _{j}\right\vert ^{2}j.  \label{8}
\end{equation}%
Magnetization $M$ is normalized in terms of $N$ and maximum of $\left\vert
M\right\vert $ is equal to 1. One has also to take into account the
normalization condition for the order parameter:

\begin{equation}
\int dS\Psi _{j}\Psi _{j}^{\ast }=1.  \label{9}
\end{equation}

The spatial profiles of all the components of the order parameter in the
equilibrium can be found from the condition of minimum of energy (4). It is
also convenient to introduce the longitudinal $l_{z}$ and transverse $l_{tr}$
local magnetizations of the condensate:

\begin{equation}
l_{z}=\left\vert \Psi _{1}\right\vert ^{2}-\left\vert \Psi _{-1}\right\vert
^{2},  \label{10}
\end{equation}

\begin{eqnarray}
l_{tr}^{2} &=&l_{x}^{2}+l_{y}^{2}  \notag \\
&&=2\left\vert \Psi _{0}\right\vert ^{2}\left\vert \Psi _{1}\right\vert
^{2}+2\left\vert \Psi _{0}\right\vert ^{2}\left\vert \Psi _{-1}\right\vert
^{2}  \notag \\
&&+4\left( \Psi _{0}^{2}\Psi _{1}^{\ast }\Psi _{-1}^{\ast }+c.c.\right) .
\label{11}
\end{eqnarray}%
The spin energy in this case can be represented as \bigskip 
\begin{equation}
F_{spin}=2\pi g_{s}\hbar \omega _{\perp }N\int dS\left(
l_{z}^{2}+l_{tr}^{2}\right) .  \label{12}
\end{equation}

In this paper, we restrict ourselves only to the case of axially-symmetric
phases, when moduli of all the components of the order parameter are
independent on the azimuthal angle and depend only on the radial coordinate $%
r$. Note that equilibrium vortex phases in this situation were studied in
Refs. \cite{Mizushima1,Isoshima1} for the spin $F=1$ condensate and in Ref. 
\cite{Pogosov} for $F=2$ system. For the axially-symmetric phases, each
component of the order parameter can be represented as 
\begin{equation}
\Psi _{j}\left( r,\varphi \right) =f_{j}(r)\exp (-iL_{j}\varphi -i\delta
_{j}),  \label{13}
\end{equation}%
where $\varphi $ is a polar angle, $L_{j}$\ is a winding number, and $\delta
_{j}$ is a relative phase. We will denote such phases as ($L_{-1}$, $L_{0}$, 
$L_{1}$). As it was shown in Ref. \cite{Isoshima1}, an axial symmetry of the
solution implies that winding numbers satisfy the relation: $%
L_{1}+L_{-1}=2L_{0}$. In this case, according to Eqs. (11) and (12), the
spin energy depends on relative angle $\chi =2\delta _{0}-\delta _{1}-\delta
_{-1}.$

It is important to note that only a spin contribution to the total energy
(4) depends on phases $\delta _{j}$ via the spin-mixing term. For the
stationary state, which is a local minimum of Gross-Pitaevskii functional
(4), the value of $\chi $ is determined by the sign of interaction constant
in a spin channel $g_{s}$. For positive $g_{s}$ (antiferromagnetic case), a
minimum of $F_{spin}$ is attained at $\chi =\pi $, whereas for negative $%
g_{s}$ (ferromagnetic case) $\chi =0$.

\section{Results and discussion}

According to the results of Ref. \cite{Isoshima1}, for the antiferromagnetic
state ($g_{s}>0$), phases (-1, 0, 1) and (1, 1, 1) are energetically
favorable in the region of small and moderate values of magnetization $M$.
Phase (-1, 0, 1) is realized at low rotation frequencies $\Omega $ and (1,
1, 1) at higher $\Omega $. In Ref. \cite{Mizushima1}, it was shown that (0,
1, 2) state is favorable in the ferromagnetic case ($g_{s}<0$) in a region
of moderate values of $\Omega $ and $M.$ In these phases, all three hypefine
states are populated. Fluctuations of $\chi $ have a sense only in this
case, since the $\chi $-dependent part of the energy is equal to zero
identically, if one of the components of the order parameter is zero. In
this paper, we will concentrate on these three vortex states, since they are
appropriate candidates for the illustration of the effect of thermal
fluctuations. Note that in homogeneous spin-1 condensate atoms populate only
two or one hyperfine state(s); they can populate three states only if the
system is trapped and experiences rotation, which generates vortices.

An important feature of real atomic spinor Bose-Einstein condensates is a
weakness of the spin interactions comparing to the interaction in density
channel ($\left\vert g_{s}\right\vert \ll \left\vert g_{n}\right\vert $). At
the same time, the coherence among the different components of the order
parameter (angle $\chi $) is fully determined by the spin interaction. Angle 
$\chi $ also influences the transverse magnetization of the condensate, as
seen from Eq. (11). Note that a longitudinal component of magnetization is
independent on $\chi $.

Smallness of $g_{s}$ comparing to $g_{n}$ leads to the fact that thermal
fluctuations of relative angle $\chi $ become significant at much lower
temperatures than fluctuations of the density of particles. Therefore, at
relatively low temperatures, one can assume that the moduli of all the
components of the order parameter remain fixed (that can be also checked
numerically), whereas $\chi $ is fluctuating. For the case of small
fluctuations of $\chi $, one can use a harmonic approximation and represent
the deviation of the energy of the system from the equilibrium, $\delta
F=F(\chi _{0}+\delta \chi )-F(\chi _{0})$, as a quadratic function in terms
of the deviation of angle $\chi $ from the equilibrium $\delta \chi =\chi
-\chi _{0}$:\bigskip 
\begin{eqnarray}
\delta F &=&2\pi g_{s}\hbar \omega _{\perp }NI\left( \cos (\chi _{0}+\delta
\chi )-\cos (\chi _{0})\right)  \notag \\
&\approx &\pi \left\vert g_{s}\right\vert \hbar \omega _{\perp }NI\left(
\delta \chi \right) ^{2},  \label{14}
\end{eqnarray}%
where $I=\int dS\left( f_{1}f_{-1}f_{0}^{2}\right) $. Under these
assumptions, the average square of the deviation of $\chi $ from the
equilibrium is given by

\begin{equation}
\left\langle (\delta \chi )^{2}\right\rangle _{T}=\frac{\int d\left( \delta
\chi \right) (\delta \chi )^{2}\exp (-\frac{\delta F}{k_{B}T})}{\int d\left(
\delta \chi \right) \exp (-\frac{\delta F}{k_{B}T})}.  \label{15}
\end{equation}%
Integrals in Eq. (15) can be calculated analytically. After taking into
account Eq. (3), we get\bigskip :%
\begin{equation}
\left\langle (\delta \chi )^{2}\right\rangle _{T}=\frac{t}{1.28\sqrt{N}%
4\left\vert g_{s}\right\vert I}.  \label{16}
\end{equation}%
We also introduce a quantity $\Delta \chi =\sqrt{\left\langle (\delta \chi
)^{2}\right\rangle _{T}}$, which can be considered as an average deviation
of angle $\chi $ from the equilibrium. We see that $\Delta \chi $ depends on
dimensionless temperature $t=T/T_{c}$, the number of particles $N$ and
integral $I$. For a given vortex phase, $I$ is also a function of total
magnetization $M$. It is important to emphasize that the scaling relation
(16) has a sense only if $g_{n}$ is independent of $N$, as discussed above.
In order to calculate $I$, we use a variational method, which was previously
applied by us in Ref. \cite{Pogosov} to evaluate energies of various
axially-symmetric vortex phases in spin $F=2$ condensate. In this approach,
each component of the order parameter is modeled by a trial function and
values of variational parameters are found from the condition of minimum of
total energy.

In Fig. 1 we plot calculated dependence of $\Delta \chi $ (measured in
degrees) as a function of the number of particles in the system for
different vortex phases at $t=0.1$ and $g_{n}=10$. This value of $g_{n}$ is
close to typical experimental ones ($a_{0}\approx 5$ nm, $n_{z}\approx 2$ nm$%
^{-1}$), see also calculations of Ref. \cite{Pogosov,Isoshima1}. We assume
that, for (-1, 0, 1) and (1, 1, 1) states, $g_{s}=0.01g_{n}$ and $M=0.1$,
whereas for (0, 1, 2), $g_{s}=-0.01g_{n}$ and $M=0.5$. Note that $\Delta
\chi $ for the particular phase is independent on $\Omega $, since $I$ has
the same property. We see that even for quite low temperatures, $\Delta \chi 
$ can be rather large and the coherence among different components of the
order parameter is practically destroyed. For smaller value of $\left\vert
g_{s}\right\vert $, fluctuations of $\chi $ are, of course, even stronger.
To illustrate the effect of temperature, in the inset to Fig. 1, we show the
dependence of $\Delta \chi $ on $T$ for (0, 1, 2)-phase (curve 1), (1, 1, 1)
phase (curve 2), and (-1, 0, 1) phase (curve 3) at fixed number of atoms $%
N=1000$, $g_{s}=-0.05g_{n}$ for the first curve and $g_{s}=0.05g_{n}$ for
two others. Note that $\Delta \chi $ is almost independent on total
magnetization $M$ of the condensate.

As we already pointed out, fluctuations of $\chi $ lead to that of $l_{tr}$.
In a harmonic approximation, one can express the average deviation of $%
\left\vert l_{tr}\right\vert $ from the equilibrium $\left\langle \delta
\left\vert l_{tr}\right\vert \right\rangle _{T}$ through the deviation of $%
\chi $:

\begin{equation}
\frac{\left\langle \delta \left\vert l_{tr}\right\vert \right\rangle _{T}}{%
\left\vert l_{tr}\right\vert }=(-1)^{u}\frac{1}{2}\left\langle (\delta \chi
)^{2}\right\rangle _{T}\frac{f_{1}f_{-1}}{(f_{1}+(-1)^{u+1}f_{-1})^{2}},
\label{17}
\end{equation}%
where $u=0$ for the antiferromagnetic case and $u=1$ for the ferromagnetic
one. If in the antiferromagnetic state the total magnetization is not large, 
$M\lesssim 0.5$, one can expect that $(f_{1}-f_{-1})^{2}\ll f_{1}f_{-1}$,
and, therefore, even small fluctuations of $\chi $ lead to strong relative
fluctuations of $\left\vert l_{tr}\right\vert $. At the same time, for
ferromagnetic case, $u=1$ in this equation, and relative fluctuations of $%
\left\vert l_{tr}\right\vert $ are much smaller.

We have calculated $\left\langle \delta \left\vert l_{tr}\right\vert
\right\rangle _{T}$ for different vortex phases and our calculations
revealed that $\left\langle \delta \left\vert l_{tr}\right\vert
\right\rangle _{T}/\left\vert l_{tr}\right\vert $ is almost independent on
radial coordinate $r$ for vortex phases (-1, 0, 1) and (1, 1, 1). This is
due to the fact that $\left\vert L_{-1}\right\vert =\left\vert
L_{1}\right\vert $ for these states, therefore, $f_{1}(r)$ is nearly
proportional to $f_{-1}(r)$, and, according to Eq. (15), $\left\langle
\delta \left\vert l_{tr}\right\vert \right\rangle _{T}/\left\vert
l_{tr}\right\vert $ should only slightly depend on $r$. In Fig. 2 we present 
$\left\langle \delta \left\vert l_{tr}\right\vert \right\rangle
_{T}/\left\vert l_{tr}\right\vert $ as a function of total magnetization of
the condensate for (-1, 0, 1) state at $t=0.1$, $g_{s}=0.01g_{n}$
(antiferromagnetic case), and $N=1000$. We see that relative fluctuations of
transverse magnetization can be significant even at low temperature. Value
of $\left\langle \delta \left\vert l_{tr}\right\vert \right\rangle
_{T}/\left\vert l_{tr}\right\vert $ decreases with increase of $M$. This
result is natural, since condensate becomes more polarized with growing $M$.
An absolute value of $\left\langle \delta \left\vert l_{tr}\right\vert
\right\rangle _{T}$ also remains sizable. Although value of fractional
quantity $\left\langle \delta \left\vert l_{tr}\right\vert \right\rangle
_{T}/\left\vert l_{tr}\right\vert $ is growing with decreasing of $M$, the
value of $\left\vert l_{tr}\right\vert $ itself becomes smaller. Therefore,
we found that the most appropriate values of $M$ to observe fluctuations of
transverse magnetization is around $M=0.2$, where both $\left\langle \delta
\left\vert l_{tr}\right\vert \right\rangle _{T}/\left\vert l_{tr}\right\vert 
$ and $\left\vert l_{tr}\right\vert $ are high: $\left\langle \delta
\left\vert l_{tr}\right\vert \right\rangle _{T}/\left\vert l_{tr}\right\vert
\gtrsim 0.1$, whereas $l_{tr}$ is comparable to the longitudinal
magnetization $l_{z}$ in the fully polarized state at $M=1$, where it should
be easily detectable experimentally. Value of $\left\langle \delta
\left\vert l_{tr}\right\vert \right\rangle _{T}/\left\vert l_{tr}\right\vert 
$ depends also on the vortex phase; we found that in (1, 1, 1) state it is
even much larger than in (-1 ,0 ,1) state.

Also we have calculated $\left\langle \delta \left\vert l_{tr}\right\vert
\right\rangle _{T}$ for the ferromagnetic (0, 1, 2) phase. As can be
expected, in this case, relative fluctuations of $\left\vert
l_{tr}\right\vert $ are much weaker. Physically, this is because $\left\vert
l_{tr}\right\vert $ is proportional to the ferromagnetic order parameter 
\cite{Pogosov}, which is responsible for the ferromagnetic ordering.
Therefore, one can expect that in the ferromagnetic phase this order
parameter is more robust with respect to thermal fluctuations, than in the
antiferromagnetic one. In addition, average deviation of $\left\vert
l_{tr}\right\vert $ from the equilibrium is negative and its modulus is
growing with increase of $M$, in contrast to the antiferromagnetic system.

Thermal fluctuations should also be important in the case of $F=2$
condensate, where there are two interaction constants in spin channel and
two characteristic angles. Therefore, one can expect more complicated
behavior, as compared to $F=1$ condensate. For instance, in homogeneous $F=2$
system, a cyclic state can have a lowest energy; in this case atoms populate
three hyperfine states, and the spin energy depends on the coherence among
them. Fluctuation problem for this system was analyzed in Ref. \cite{Pogo}.
A new method to create such entangled states in spin-1 condensate was
recently applied experimentally in Ref. \cite{Chapman}, where a microwave
energy was injected to the system. As a result, particles redistribute from
spin $-1$ state to spin $0$ and $1$ states, and all three magnetic sublevels
become populated. The spin-mixing dynamics in $F=1$ condensate was studied
theoretically in Ref. \cite{Pu}.

Note that in Eq. (14) we have assumed that fluctuating $\chi $ is spatially
independent that is not true in general case. However, spatial gradients of $%
\chi $ give some additional contribution to the kinetic energy of the
system, which is much larger than the spin energy. Therefore, gradients of $%
\chi $\ result in rather large increase of total energy, and we can neglect
them for the trapped system, at least for our range of parameters. In other
words, healing length for $\chi $ far exceeds the Thomas-Fermi radius of the
system, and, therefore, although $\chi $ is fluctuating inside the cloud, it
remains nearly constant \cite{Pogo}, except of the surface layer, where the
density of particles is low.

Thermal fluctuations of $\chi $ should be also noticeable in
three-dimensional condensates at low and moderate temperatures. In general,
the dependences of the number of condensed particles on the reduced
temperature and critical temperature on the total number of atoms for 3D
case are similar to that in 2D system, which are described by Eqs. (2) and
(3). The main difference is the powers of $t$ and $N$ in the right hand
sides of Eqs. (2) and (3) that are $3$ and $-1/3$ ($\hbar \omega _{\perp
}/kT_{c}\sim N^{-1/3}$), respectively. However, in this case one has to take
accurately into account the possibility of long wave length fluctuations of $%
\chi $ in longitudinal direction and formation of kinks \cite{Pogo}.

\section{Conclusions}

In this paper, we have studied the effect of thermal fluctuations on the
coherence among different components of the order parameter in quasi 2D
rotating $F=1$ Bose-Einstein condensate, when all three hyperfine states are
populated. Different axially-symmetric vortex phases were considered. We
have shown that the deviation of the relative phase $\chi =2\delta
_{0}-\delta _{1}-\delta _{-1}$ from the equilibrium can be very significant
even at low temperatures, much smaller than $T_{c}$. Fluctuations of
relative angle induce sizable fluctuations of the spin texture, namely,
local transverse magnetization of the condensate. We have shown that these
fluctuations are much more pronounced in antiferromagnetic case than in the
ferromagnetic one. The recently proposed in Ref. \cite{Higbie} direct and
nondestuctive method for the imaging of spinor BEC spatial magnetization (or
some of its modification) can be applied for the experimental study of the
thermal fluctuations of spin textures, since it enables multiple-shot
imaging and one can directly observe the dynamics of a single sample.

\bigskip \acknowledgments

Authors acknowledge useful discussions with T. K. Ghosh and T. Mizushima. W.
V. Pogosov is supported by the Japan Society for the Promotion of Science.

\section{Figure captions}

\textbf{Fig. 1.} Dependences of $\Delta \chi $ (in degrees) on the number of
particles in the system for different vortex phases at fixed value of
interaction constant $g_{n}=10$ (see in the text) and $t=0.1$. In the (-1,
0, 1) phase, $g_{s}=0.01g_{n}$, $M=0.1$; in the (1, 1, 1) state, $%
g_{s}=0.01g_{n}$, $M=0.1$; in the (0, 1, 2) state, $g_{s}=-0.01g_{n}$, $%
M=0.5 $. Inset shows $\Delta \chi $ as a function of temperature for (0, 1,
2) phase (curve 1), (1, 1, 1) phase (curve 2), and (-1, 0, 1) phase (curve
3) at the same values of $M$, $g_{n}$\ and $t$. Number of atoms is $N=1000$,
interaction constants are $g_{s}=-0.05g_{n}$ for the first curve and $%
g_{s}=0.05g_{n}$ for two others.

\textbf{Fig. 2.} Dependences of $\left\langle \delta \left\vert
l_{tr}\right\vert \right\rangle _{T}/\left\vert l_{tr}\right\vert $ on the
total magnetization for (-1, 0, 1) phase at $N=1000$; $g_{n}=10$, $t=0.1$, $%
g_{s}=0.01g_{n}$.

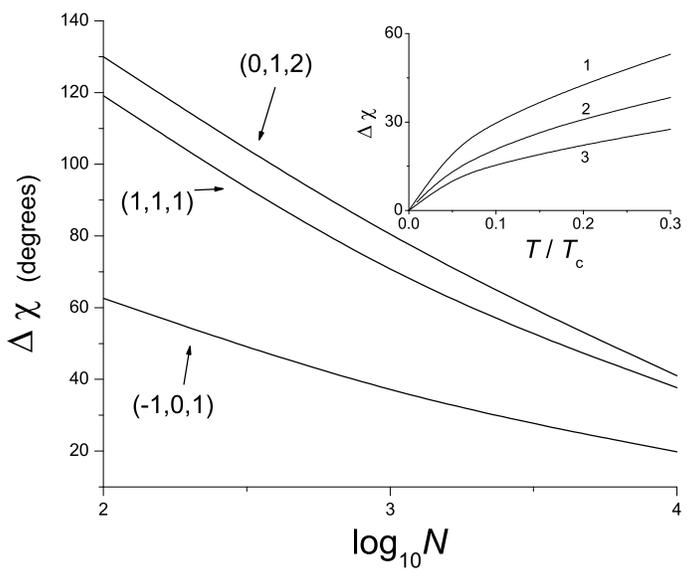
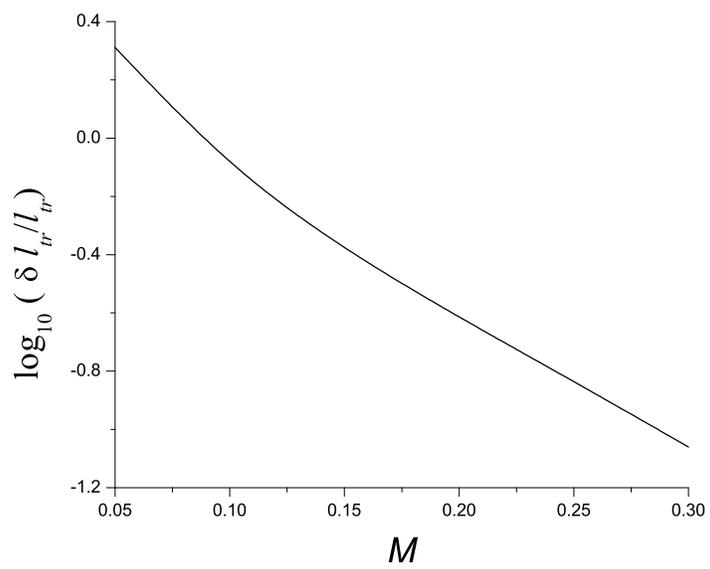
\end{document}